\documentclass[aps,pra,twocolumn,superscriptaddress,amsmath,amssymb]{revtex4}
\usepackage{graphicx}
\usepackage{epstopdf}
\usepackage{amsmath}
\usepackage{color}
\usepackage[normalem]{ulem}
\usepackage{graphicx}
\usepackage{subfigure}
\usepackage{float}
\usepackage{mathrsfs}
\usepackage{bbold}
\usepackage{psfrag}
\usepackage{mathcomp}
\usepackage{verbatim}
\usepackage{multirow}
\usepackage{diagbox}
\usepackage{slashed}
\usepackage{braket}
\usepackage{leftidx}
\usepackage[colorlinks,citecolor=blue]{hyperref}
\usepackage[toc,page,title,titletoc,header]{appendix}

\begin{document}

\title{Interaction induced dynamical $\mathcal{PT}$  symmetry breaking in dissipative Fermi-Hubbard models}
\author{Lei Pan}
\affiliation{Beijing National Laboratory for Condensed Matter Physics, Institute of
Physics, Chinese Academy of Sciences, Beijing 100190, China}
\affiliation{Institute for Advanced Study, Tsinghua University, Beijing, 100084, China}
\author{Xueliang Wang}
\affiliation{Beijing National Laboratory for Condensed Matter Physics, Institute of
Physics, Chinese Academy of Sciences, Beijing 100190, China}
\affiliation{School of Physical Sciences, University of Chinese Academy of Sciences, Beijing 100049, China}
\author{Xiaoling Cui}
\affiliation{Beijing National Laboratory for Condensed Matter Physics, Institute of Physics, Chinese Academy of Sciences, Beijing 100190, China}
\affiliation{Songshan Lake Materials Laboratory , Dongguan, Guangdong 523808, China}
\author{Shu Chen}
\email{schen@iphy.ac.cn}
\affiliation{Beijing National Laboratory for Condensed Matter Physics, Institute of Physics, Chinese Academy of Sciences, Beijing 100190, China}
\affiliation{School of Physical Sciences, University of Chinese Academy of Sciences, Beijing 100049, China}
\affiliation{Yangtze River Delta Physics Research Center, Liyang, Jiangsu 213300, China}
\date{\today}

\begin{abstract}
We investigate the dynamical properties of one-dimensional dissipative Fermi-Hubbard models, which are described by the Lindblad master equations with site-dependent jump operators. 
The corresponding non-Hermitian effective Hamiltonians with pure loss terms possess parity-time ($\mathcal{PT}$)
symmetry if we compensate the system additionally an overall gain term.
By solving the two-site Lindblad equation with fixed dissipation exactly, we find that the dynamics of rescaled density matrix shows an instability as the interaction increases over a threshold, which can be equivalently described in the scheme of  non-Hermitian effective Hamiltonians. This instability is also observed in multi-site systems and  closely related to the $\mathcal{PT}$ symmetry breaking accompanied by appearance of complex eigenvalues of the effective Hamiltonian.
Moreover, we unveil that the dynamical instability of the anti-ferromagnetic Mott phase comes from the $\mathcal{PT}$ symmetry breaking in highly excited bands, although the low-energy effective model of the non-Hermitian Hubbard model in the strongly interacting regime is always Hermitian.
We also provide a quantitative estimation of the time for the observation of dynamical $\mathcal{PT}$ symmetry breaking which could be probed in experiments.

\end{abstract}
\maketitle
\section{Introduction}\label{sec1}
Dissipations are ubiquitous in nature since the physical systems are always inevitable to be influenced by surroundings.
Under Markov approximation, the dynamics of dissipative quantum system is commonly described by Lindblad
master equation \cite{Markov}. Since solving the Lindblad equation accurately is quite expensive, an approximate but efficient approach to describe the time evolution of open quantum system is directly handling the stationary Schr\"{o}dinger equation, in which case the dynamics is determined by an effective non-Hermitian Hamiltonian under certain condition \cite{Daley,Castin}. Under this scheme, it is feasible to study the dissipative dynamics by means of the spectrum of non-Hermitian Hamiltonian.
Among various kinds of non-Hermitian Hamiltonians, there exists a fascinating one, i.e., the parity-time ($\mathcal{PT}$) symmetric Hamiltonian $\left[\mathcal{PT},H\right]=0$ whose spectra can be real and bounded below \cite{Bender1998}.
There is a notable feature that the $\mathcal{PT}$ symmetry of a system can be spontaneously broken and the system undergoes a transition from $\mathcal{PT}$ unbroken phase to a spontaneously $\mathcal{PT}$ broken phase. It is shown that the eigenvalues in the $\mathcal{PT}$ unbroken phase are always purely real, while complex conjugated eigenvalues appear in the spectrum for the $\mathcal{PT}$ broken phase and the system exhibits many novel phenomena once $\mathcal{PT}$ symmetry broken transition happens.

During the past decade, $\mathcal{PT}$-symmetric systems and the corresponding $\mathcal{PT}$ symmetry breaking have been experimentally realized and explored in various classical systems \cite{Guo,Longhi1,Feng1,Ruter,Lin1,Regensburger,Feng2,Feng3,Hodaei1,Yang3,Hodaei2,Yang4,Musslimani,Fatkhulla,Andrey,Wang,Wimmer,Schindler,ZhangX,Cummer,Ding2,YangL,ChangL}
and quantum systems including quantum gas \cite{LuoL}, single spin system \cite{DuJ}, synthetic lattice \cite{Lapp} and photonic systems \cite{XueP,Klauck}.
Whereas the above experiments on quantum $\mathcal{PT}$ systems mainly focus on single particle physics \cite{LuoL,DuJ,Lapp,XueP,Klauck},
recent theoretical studies have revealed intriguing physical properties with the interplay of the non-Hermitian effects and interactions \cite{Pan1,Ueda1,Cui,Ueda4,Ueda5,YuZ}, such as enhanced sensitivity at exceptional points \cite{Pan1,YuZ}, non-Hermitian superfluidity \cite{Ueda1} and enhanced pairing superfluidity \cite{Cui}, $\mathcal{PT}$ symmetric quantum critical phenomena \cite{Ueda4,Ueda5}, anomalous slow dynamics in quantum criticality \cite{ZhaiH}, and nontrivial non-Hermitian many-body topological phases \cite{NH-FQH,XuZH,ZhuSL,Nori}.
Particularly, several recent works reported experimental studies of dissipative many-body systems with controllable dissipation in bosonic optical lattices \cite{Bouganne,Tomita1,Tomita2}, which have stimulated theoretical interests in exploring novel physical phenomena induced by the interplay between interaction and dissipation \cite{ZCai,Daley,Pan1,Ueda1,Cui,Ueda4,Ueda5,YuZ,ZhaiH,NH-FQH,XuZH,PRResearch,ZhuSL,Nori}.

\begin{figure}[htb]
\centering
\includegraphics[width=7.5cm]{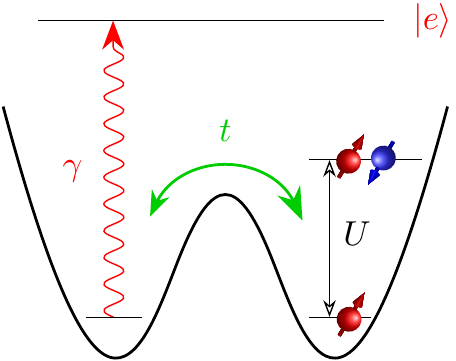}
\caption{Schematic of the experimental setup for double-well Fermi-Hubbard model with a site-dependent  dissipation. A resonant laser is used to create a dissipation with strength $\gamma$ on the left well.}
\label{Fig1}
\end{figure}

In this work, we study the effect of interaction on the quantum dynamics in 1D dissipative Fermi-Hubbard models with effective parity-time symmetry. To begin with, we consider the two-site system with site-dependent dissipation, which is created by a laser beam acting on one of lattice sites to ensure the emergence of  $\mathcal{PT}$ symmetry, as schematically shown in Fig.\ref{Fig1}. By numerically solving the Lindblad master equation for this double-well model, we find that the dissipative system can exhibit a clear dynamical signature of  $\mathcal{PT}$ symmetry breaking characterized by an anomalous dynamical instability, if we rescale the density operator by multiplying an overall exponential factor. 
We further reveal the important role of interaction and demonstrate that a strong interaction can induce dynamical instability even in the low dissipation regime, which can be alternatively understood by the effective non-Hermitian  Fermi Hubbard model with site-dependent imaginary potentials. 
In the strong interaction limit, the low-energy physics of half-filled fermions in the non-Hermitian dissipative lattice can be 
effectively described by a Hermitian anti-ferromagnetic (AFM) spin exchange model. However, the fermion dynamics is still unstable due to the existence of complex eigenvalues for the highly excited states. This instability persists in multi-site systems and leads to the dynamical instability of Mott phase. 
We estimate the lifetime of AFM state in Mott phase by degenerate perturbation analysis.
Our results demonstrate that the interplay between interaction and dissipation leads to dynamical  $\mathcal{PT}$ symmetry breaking and unstable quantum dynamics in the dissipative Fermi-Hubbard model, which are very different dynamical behaviors from its Hermitian correspondence.

The rest of the paper is organized as follows: In Sec. II
we present the formalism of solving the dynamics of two-site dissipative Hubbard model by using both the Lindblad master equation and effective non-Hermitian Hamiltonian.  We demonstrate that the two methods play an equivalent role in dealing with the dynamical problem for the initial state with fixed particle number. In
Sec. III, we study the multi-site systems in the scheme of effective non-Hermitian Hamiltonian. Finally, we
conclude and discuss the potential experimental detection in Sec. IV.
\section{Dissipative Fermi-Hubbard model: Two-site system} \label{sec2}
To investigate the quantum dynamics of an open quantum system with dissipation, we take advantage of the Lindblad master equation ($\hbar=1$) \cite{Lindblad,Lindblad2}
\begin{eqnarray}
\frac{d\varrho(\tau)}{d \tau }&=&-i[H_s,\varrho]+\sum_k\Big(L_k\varrho L_k^{\dag}-\frac{1}{2}\{L_k^{\dag}L_k,\varrho\}\Big) \nonumber \\
&=&-i\big(\mathcal{H}\varrho-\varrho \mathcal{H}^{\dag}\big)+\sum_kL_k\varrho L_k^{\dag}, \label{Eq2.1}
\end{eqnarray}
where $\varrho(\tau)$ denotes the density matrix, $H_s$ represents Hamiltonian of the system and $L_k$ is Lindblad operator with the index $k$ (such as space, spin degrees of freedom, etc.). Here  $\mathcal{H}$  is called effective non-Hermitian Hamiltonian defined by
\begin{equation}
\mathcal{H}\equiv H_s-\frac{i}{2}\sum_kL_k^{\dag}L_k,
\end{equation}
which includes a non-Hermitian part. In this work, we shall consider the Hubbard model with $L$ sites, i.e.,
\begin{equation}
H_s = -t\sum_{j=1,\sigma}^{L-1}(c_{j\sigma}^{\dag}c_{j+1,\sigma}+h.c)+U \sum_{j=1}^{L} n_{j\uparrow}n_{j\downarrow}, \label{Hubbard}
\end{equation}
where the symbol $j$ ($\sigma$) denotes the site (spin) index, $t$ represents the hopping term and $U$ the interaction strength. For convenience, we shall set $t=1$ as the energy unit throughout.

In order to present the formalism clearly and get an intuitive understanding, we firstly focus on the double well system and study its dynamics in details. The Lindblad operators are chosen as $L_{k=(1,\uparrow)}=2\sqrt{\gamma} c_{1\uparrow}, L_{k=(1,\downarrow)}=2\sqrt{\gamma} c_{1\downarrow}$, which  represents that a single-particle dissipation is engineered on the left well with strength $\gamma$ (see Fig.\ref{Fig1}), and then the Lindblad master equation becomes
\begin{eqnarray}
\frac{d\varrho(\tau)}{d \tau}=-i\big(\mathcal{H}\varrho-\varrho \mathcal{H}^{\dag}\big)
+4\gamma c_{1\uparrow}\varrho c_{1\uparrow}^{\dag}+4\gamma c_{1\downarrow}\varrho c_{1\downarrow}^{\dag}.
\label{Eq2.2}
\end{eqnarray}
The effective non-Hermitian Hamiltonian is given by
\begin{eqnarray}
\mathcal{H}\equiv -t\sum_{\sigma}(c_{1\sigma}^{\dag}c_{2\sigma}+c_{2\sigma}^{\dag}c_{1\sigma})
+U\sum_{j=1}^{2}n_{j\uparrow}n_{j\downarrow}-2i\gamma n_1. \label{Eq2.3}
\end{eqnarray}
The full dynamics can be obtained once the Lindblad master equation (\ref{Eq2.2}) is solved. Since there is only loss term, the effective Hamiltonian does not posses $\mathcal{PT}$ symmetry and its spectrum consists of a series of complex eigenvalues. Nevertheless, the Hamiltonian is associated with a Hamiltonian with $\mathcal{PT}$ symmetry apart from a total damping term, which can be effectively obtained by  a dynamical rescaling of the density operator.
For convenience, we define a rescaled density matrix as
\begin{eqnarray}
\widetilde{\rho}(\tau)\equiv e^{2N\gamma \tau }\varrho_N( \tau ), \label{2.4}
\end{eqnarray}
where $N=n_1+n_2$ is the total particle number in the initial state and $\varrho_N(\tau)=P_N\varrho(
\tau) P_N$ ($P_N$ is the projection operator on $N$-particle subspace).
\begin{figure}[H]
\centering
\includegraphics[width=8.5cm]{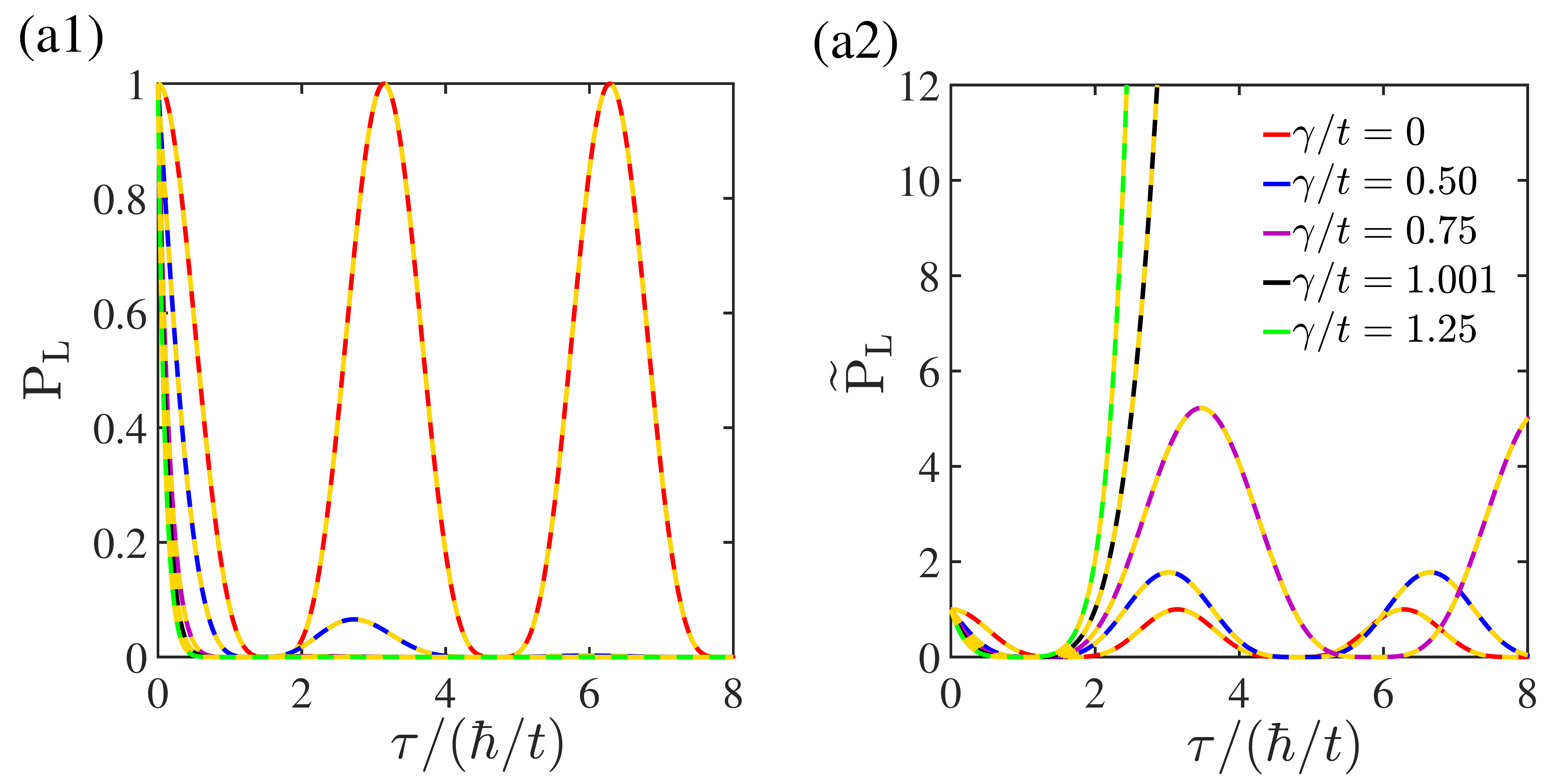}
\includegraphics[width=8.5cm]{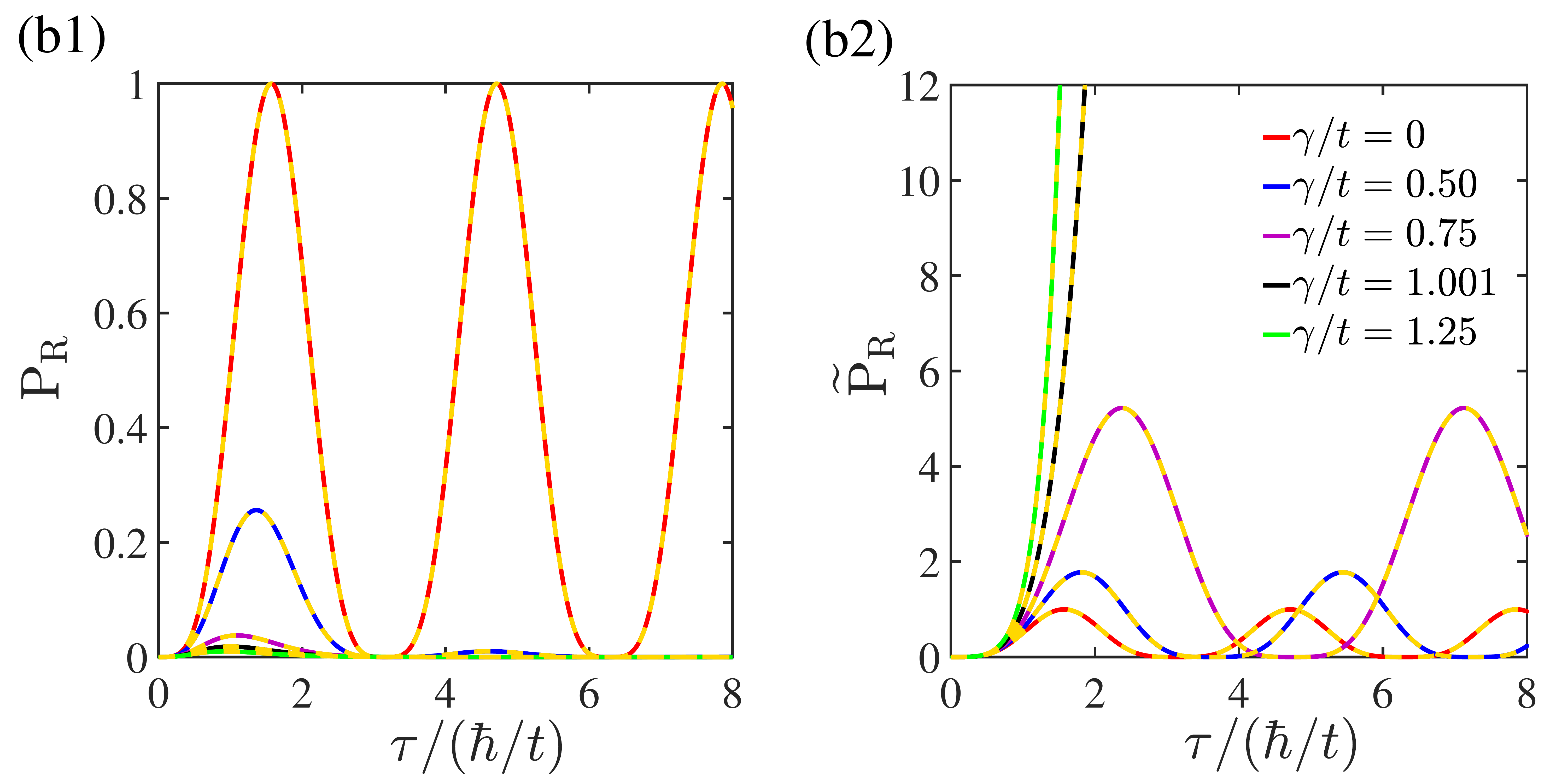}
\caption{Time evolutions of the probabilities, shown in (a1) and (b1), and rescaled probabilities, in (a2) and (b2), for both of two fermions occupying the left well and right well with dissipation strengths $\gamma/t=0$, $\gamma/t=0.25$, $\gamma/t=0.5$, $\gamma/t=0.75$, $\gamma/t=1.0$, $\gamma/t=1.25$, respectively. Here $U=0$ and the initial value is chosen by $P_L(0)=1$. The solid lines are from Lindblad master equation and dashed lines are obtained from the non-Hermitian effective Hamiltonian. Here the unit of time is $\hbar/t$.}
\label{Fig2}
\end{figure}

To solve the dynamical problem, one needs to provide an initial state since Eq.(\ref{Eq2.2}) is the first-order differential equation.
Here we choose $\varrho(0)=\ket{\uparrow\downarrow}_1\ket{0}_2 \sideset{_2}{}{\mathop{\bra{0}}}\sideset{_1}{}{\mathop{\bra{\downarrow\uparrow}}}$ as the density matrix for the initial state with both the fermions occupying the left well.
We firstly discuss the non-interaction case ($U=0$). In Fig.\ref{Fig2} (a1) and (a2), we display the survival probability and rescaled survival probability of two fermions in the left well, given by $P_L(\tau)=\sideset{_2}{}{\mathop{\bra{0}}}\leftidx{_1}{\bra{\downarrow\uparrow}}\rho(\tau)\ket{\uparrow\downarrow}_1\ket{0}_2 $ and $\widetilde{P}_L(\tau)=\sideset{_2}{}{\mathop{\bra{0}}}\leftidx{_1}{\bra{\downarrow\uparrow}}\widetilde{\rho}(\tau)\ket{\uparrow\downarrow}_1\ket{0}_2 $, respectively.
After some straightforward calculations, we can get
\begin{equation}
P_L(\tau)= e^{-4 \gamma \tau} f(\tau),  ~~~~\widetilde{P}_L(\tau)= f(\tau),
\end{equation}
where
\begin{equation}
f(\tau) =\left| \frac{-t^2 +(2\omega^2+t^2) \cosh(2 \omega \tau) -2 \gamma \omega \sinh(2 \omega \tau)}{2 \omega^2}\right|^2 \label{Eq8}
\end{equation}
with $\omega=\sqrt{\gamma^2-t^2}$. 
It is clear that $P_L(\tau)$ always decays with the growth of dissipation strength $\gamma$. However, the rescaled $\widetilde{P}_L(\tau)$ displays divergence behavior for $\gamma/t > 1$, whereas it is an oscillating function of time for $\gamma/t<1$.
In Fig.\ref{Fig2} (b1) and (b2), we display the time evolution of probability and rescaled probability for two fermions occupying the right well, given by  $P_R(\tau)=\sideset{_2}{}{\mathop{\bra{\downarrow\uparrow}}}\sideset{_1}{}{\mathop{\bra{0}}}\rho(\tau)\ket{0}_1\ket{\uparrow\downarrow}_2 $
and $\widetilde{P}_R(\tau)=\sideset{_2}{}{\mathop{\bra{\downarrow\uparrow}}}\sideset{_1}{}{\mathop{\bra{0}}}\widetilde{\rho}(\tau)\ket{0}_1\ket{\uparrow\downarrow}_2 $, respectively. While $P_R(\tau)$ decays very quickly as shown in Fig.\ref{Fig2} (b1),  we can see the rescaled amplitude of $\widetilde{P}_R(\tau)$ grows with the increase of $\gamma$ and then diverges once the $\gamma$ crosses a certain threshold ($\gamma_c/t=1$) as shown in Fig.\ref{Fig2} (b2).

\begin{figure}[htb]
\centering
\includegraphics[width=8.5cm]{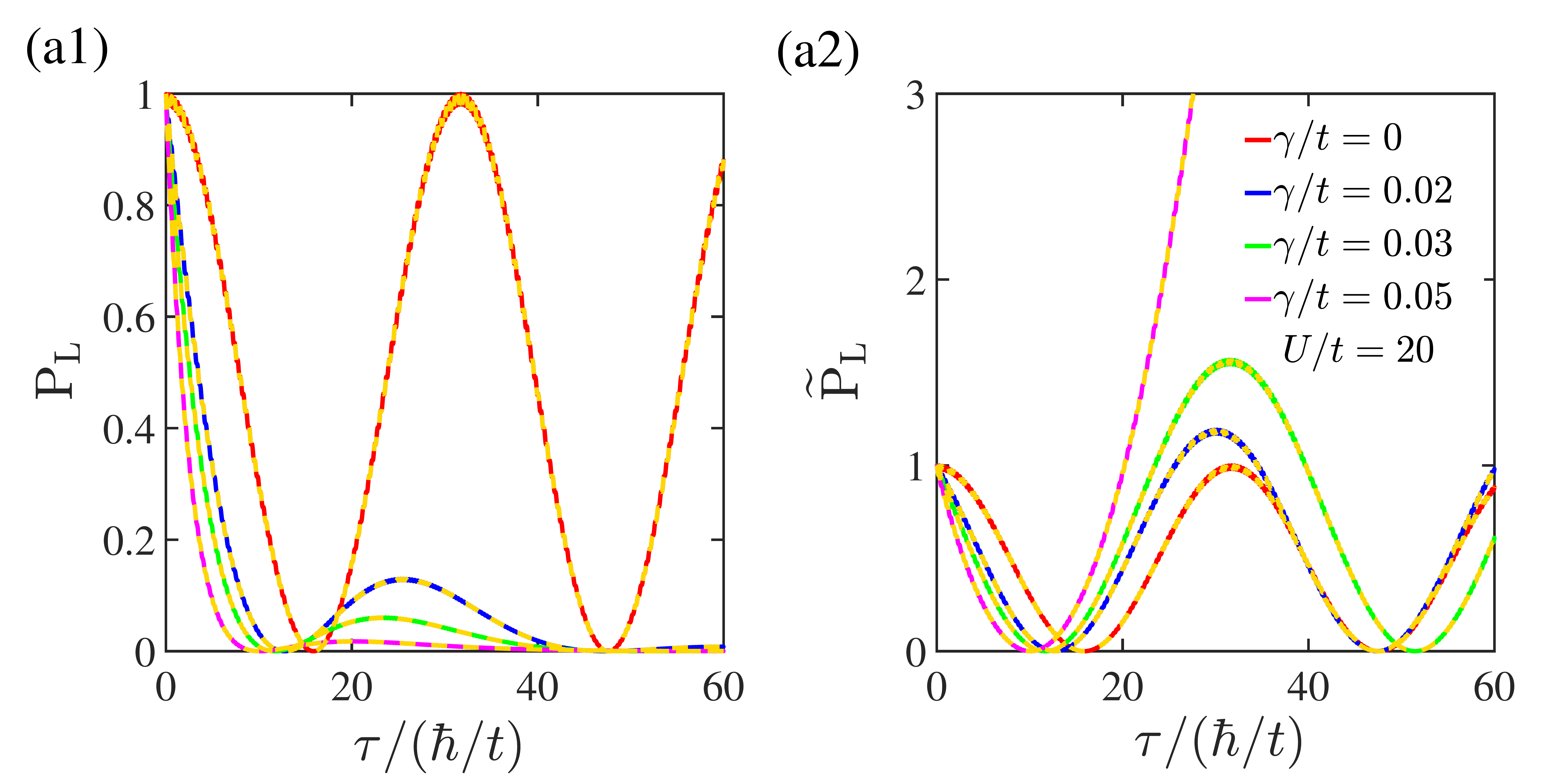}
\includegraphics[width=8.5cm]{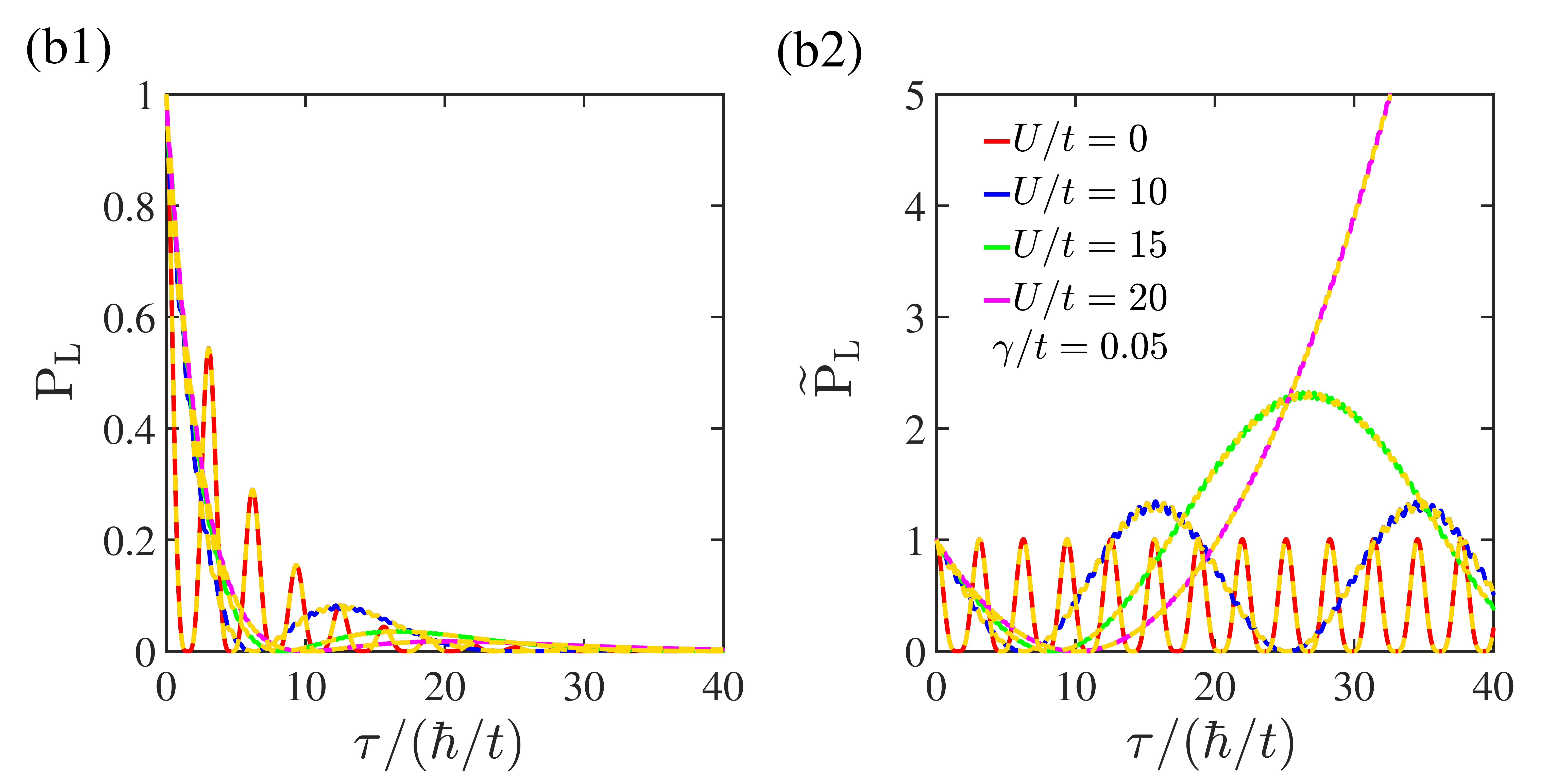}
\caption{
Time evolutions of the probabilities (a1) and rescaled probabilities (a2) of both two fermions occupying the left-well for different dissipation strengths $\gamma$ with fixed interaction strengths $U/t=20$.
Time evolutions of the probabilities (b1) and rescaled probabilities (b2) of both two fermions occupying the left-well for different interaction strengths at $\gamma/t=0.05$.
The solid lines are from Lindblad master equation and dashed lines are obtained from the non-Hermitian effective Hamiltonian. Here the unit of time is $\hbar/t$.}
\label{Fig3}
\end{figure}

Now we study the effect of interaction on the quantum dynamics and consider the case with both two fermions occupying the left well as the initial state.  In the absence of dissipation, the increase of interaction strength shall suppress the hopping of the fermion pair to its neighboring site and lead to the formation of repulsively bound pair \cite{Winkler,WangLi}, which is dynamically more stable with stronger interaction. In order to reveal the interplay between the interaction and dissipation, we display $P_L(\tau)$ and the rescaled $\widetilde{P}_L(\tau)$  with a fixed $U/t=20$ and various $\gamma/t$ in Fig.\ref{Fig3} (a1) and (a2),  and a fixed dissipative strength $\gamma/t=0.05$ and different values $U/t$ in Fig.\ref{Fig3} (b1) and (b2), respectively.
As seen in the Fig.\ref{Fig3} (a2), the rescaled $\widetilde{P}_L(\tau)$ oscillates periodically over time for $\gamma/t=0$, $0.02$, and $0.03$, but a divergence appears when  $\gamma/t=0.05$. Similarly,  the rescaled $\widetilde{P}_L(\tau)$ is divergent when the interaction strength exceeds a threshold ($U_c/t \approx 19.9$), as shown in Fig.\ref{Fig3} (b2). The divergence in the rescaled $\widetilde{P}_L(\tau)$ indicates the emergence of dynamical instability as an interplay effect of interaction and dissipation.
As we will show below, this interaction induced dynamical instability stems from the emergence of imaginary parts in effective non-Hermitian Hamiltonian which is closely related to spontaneously $\mathcal{PT}$ symmetry breaking.

To get a clear understanding, we need to unveil the equivalence in dynamics between the non-Hermitian effective Hamiltonian and Lindblad master equation.
Given the reality in experiment that the particle number $N$ of an initial state is a certain number, the initial density matrix $\varrho(0)$ is block diagonalized in particle number space,
which results in the quantum jump term $\gamma c_{j\sigma}\varrho(0)c_{j\sigma}^{\dag}$ having no effects on the dynamics in the subspace of initial particle number.
More specifically, suppose the initial density operator given by $\varrho(0)=|\psi_0(N)\rangle \langle \psi_0(N)|$ with conserving particle number $N$, the quantum jump term acts on the density matrix as $c_{j\sigma}|\psi_0(N)\rangle \langle \psi_0(N)|c_{j\sigma}^{\dag}$ and one can find the matrix elements are always zero in the subspace with the initial particle number, i.e. $P_Nc_{j\sigma}\varrho c_{j\sigma}^{\dag}P_N\equiv0$. 
In other words, the dynamics in the $N$-particle subspace is completely determined by non-Hermitian effective Hamiltonian
\begin{eqnarray}
\varrho_N(\tau)=e^{-i\mathcal{H} \tau}\varrho_N(0)e^{i\mathcal{H}^{\dagger} \tau}, \label{2.5}
\end{eqnarray}
where $\varrho_N(\tau)=P_N\varrho P_N$ and $[\mathcal{H},P_N]=0$ is also considered.
Thus, from viewpoint of the effective non-Hermitian Hamiltonian, the full dynamics of $\rho_N(\tau)$ in Eq.(\ref{2.4}) is obtained after solving Eq.(\ref{2.5}). This demonstration is also applied to other non-Hermitian systems\cite{Wunner1,Wunner2,Wunner3,Wunner4,Wunner5,ZhuBG}.
The time evolution of $\rho(\tau)$ obtained by diagonalizing Eq.(\ref{2.5}) ($\mathcal{H}$ and $\mathcal{H}^{\dagger}$) is completely consistent with the solution of the Lindblad master equation. Our numerical results also demonstrate the equivalence between the Lindblad master equation and the formalism in terms of effective non-Hermitian Hamiltonian in dealing with our studied dynamical problem as shown in Fig.\ref{Fig2} and Fig.\ref{Fig3}. Here the time evolution in effective Hamiltonian is obtained by exact diagonalization and the Lindblad master equation is solved by fourth order Runge-Kutta method. 

\begin{figure}[htb]
\centering
\includegraphics[width=9cm]{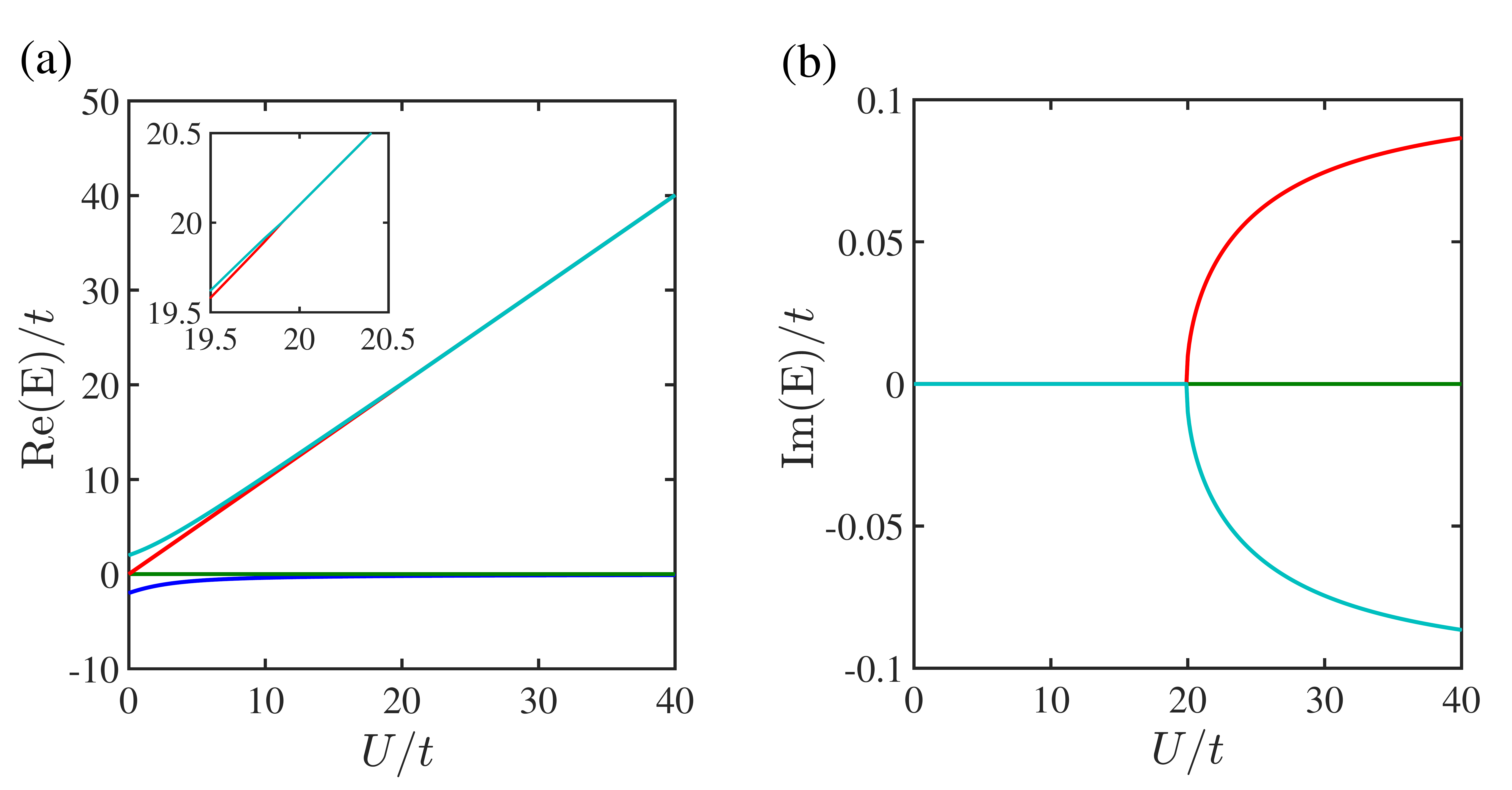}
\caption{Exact energy spectrum of $\mathcal{PT}$-symmetric double-well Fermi-Hubbard model with a spin-up and spin-down fermions. Real parts (a) and imaginary parts (b) as functions of interaction strength $U/t$. Here $\gamma/t=0.05$ and $t=1$ as the energy unit.}
\label{Fig4}
\end{figure}

In order to reveal the origin of the instability in dynamics (see Fig.\ref{Fig2} and Fig.\ref{Fig3}) from the Lindblad master equation, we rewrite the effective non-Hermitian Hamiltonian as $\mathcal{H}=H_\mathcal{PT}-iN\gamma$, then
\begin{equation}
\begin{split}
H_\mathcal{PT}&=-t\sum_{\sigma}(c_{1\sigma}^{\dag}c_{2\sigma}+c_{2\sigma}^{\dag}c_{1\sigma})
+U\sum_{j=1}^{2}n_{j\uparrow}n_{j\downarrow}\\
&-i\gamma(n_1-n_2). \label{2.6}
\end{split}
\end{equation}
One can see clearly that $H_\mathcal{PT}$ is $\mathcal{PT}$ symmetric, namely $\left[\mathcal{PT},H\right]=0$ since $\mathcal{P}c_{1(2)}\mathcal{P}=c_{2(1)}$ and $\mathcal{T}i\mathcal{T}=-i$ and then we find
\begin{eqnarray}
\widetilde{\rho} (\tau)=e^{-iH_\mathcal{PT}\tau}\varrho_N(0)e^{iH_\mathcal{PT}^{\dagger}\tau}, \label{2.7}
\end{eqnarray}
which indicates the dynamics of rescaled density matrix $\widetilde{\rho} (\tau)$ is determined by the $\mathcal{PT}$ symmetric Hamiltonian and all the information are contained in the eigenvalues and eigenvectors of $H_\mathcal{PT}$ ($H_\mathcal{PT}^\dagger$). We can deduce that the anomalous dynamical divergence is driven by the $\mathcal{PT}$ symmetry breaking since the imaginary parts of spectrum of $H_\mathcal{PT}$ in  $\mathcal{PT}$-symmetry-breaking phase always appear in pairs with complex conjugate values\cite{Bender2}.
In Fig.\ref{Fig4}, we display the energy spectrum of $H_\mathcal{PT}$ with $\gamma/t=0.05$ taken the same value as in Fig.\ref{Fig3} (b). It is clear that the $\mathcal{PT}$ symmetry is broken when interaction strength exceeds a critical value $U_c/t \approx 19.9 $. 
The dynamical instability appeared in Fig.\ref{Fig3} (b2) is induced by the $\mathcal{PT}$-symmetry breaking.

\begin{figure}[htb]
\centering
\includegraphics[width=8.0cm]{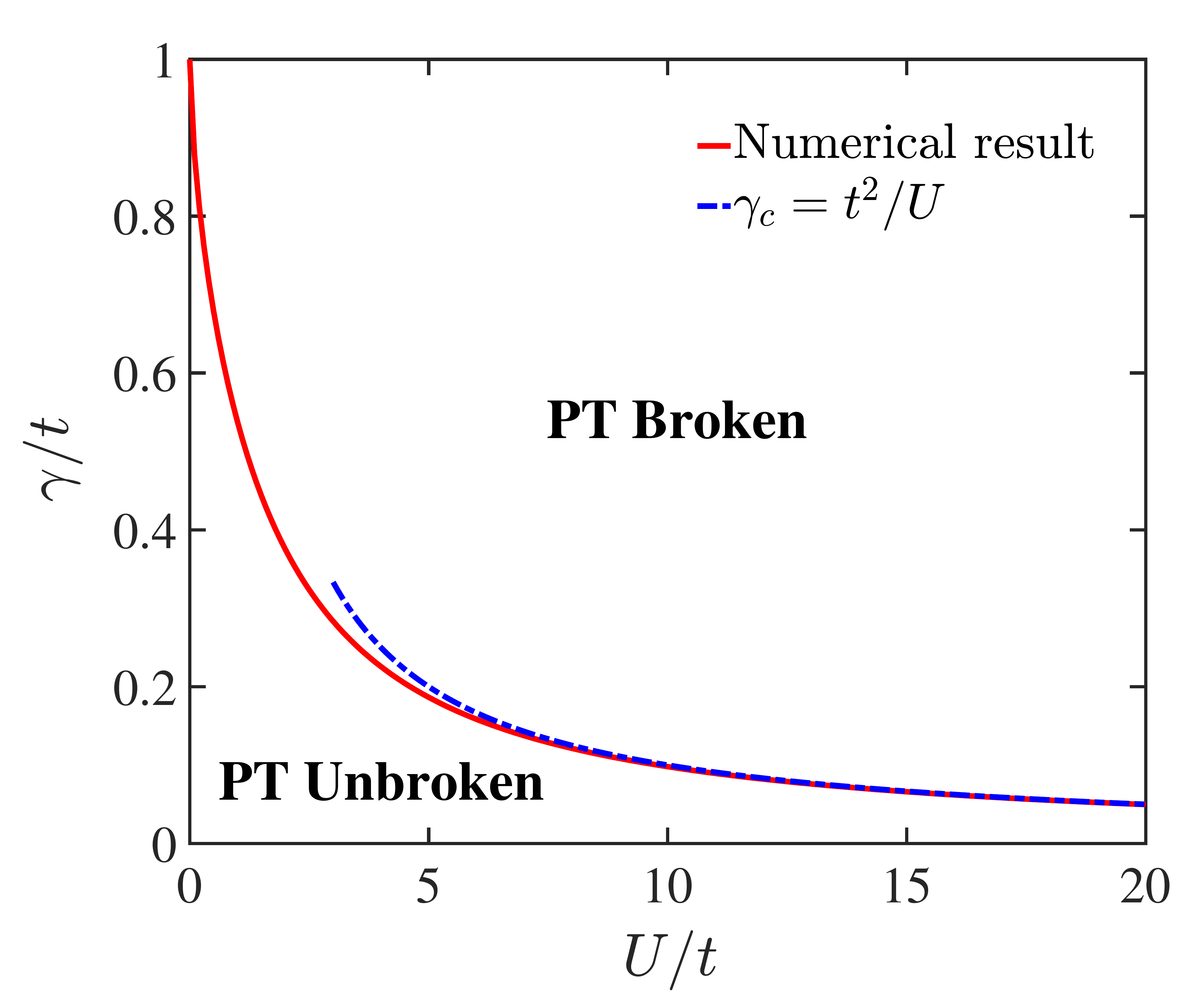}
\caption{Phase diagram of $\mathcal{PT}$-symmetric double-well Fermi-Hubbard model in $\gamma-U$ plane. The red solid line is the critical line below (above) which the system is in $\mathcal{PT}$ unbroken (broken) phase phase. The blue dash-dot line is the prediction by degenerate perturbation theory in strong interaction limit.}
\label{Fig5}
\end{figure}

In Fig.\ref{Fig5}, we plot the phase diagram in $\gamma-U$ plane for the system including one spin-up and one spin-down fermions. For a given $U$, increasing the dissipative strength $\gamma$ shall induce transition from $\mathcal{PT}$ unbroken phase to broken phase. Particularly, in the non-interacting limit $U=0$, the system is in  $\mathcal{PT}$-symmetry-breaking phase when $\gamma\geq1$, which explains the results shown in Fig.\ref{Fig2} (a2) and (b2), where the rescaled probabilities diverge once $\gamma$  is over the critical value ($\gamma_c/t=1$). On the other hand, for a fixed $\gamma/t$, the system undergoes $\mathcal{PT}$ phase transition as the interaction strength increase over a critical value.

In strong interaction regime ($U\gg t,\gamma$), we can derive the low-energy effective Hamiltonian of Eq.(\ref{2.6}), 
which reads as
\begin{eqnarray}
H_{\rm{eff}}^{\rm{low}}=J_{\rm{eff}}\Big(\mathbf{S}_1\cdot \mathbf{S}_2-\frac{1}{4}\Big),\label{2.8}
\end{eqnarray}
where $\mathbf{S_i}=(S_i^x,S_i^y,S_i^z)$ with spin-half operators $S_i^{x,y,z}$ at $i$-{\rm th} site and $J_{\rm{eff}}=\frac{4t^2U}{U^2+4\gamma^2}$.
Apparently, the effective anti-ferromagnetic (AFM) exchange model (\ref{2.8}) is a Hermitian Hamiltonian, and therefore has real eigenvalues.
Similarly, an effective Hamiltonian in the projected high-energy double-occupied subspace can be derived by means of degenerate perturbation theory and is given by
\begin{eqnarray}
H_{\rm{eff}}^{\rm{high}}=U\mathbf{I}+2i\gamma\sigma_z+\frac{2t^2}{U}\sigma_x, \label{2.9}
\end{eqnarray}
where $\mathbf{I}$ denotes the identity matrix and $\sigma_x,\sigma_z$ are Pauli matrices where spin-up (spin-down) represents
two fermions occupying the right (left) well.
Interestingly, this simple $2\times2$ Hamiltonian predicts the $\mathcal{PT}$-symmetry broken transition when interaction crosses the critical point $U_c=\frac{t^2}{\gamma}$, which agrees with the numerical result very well in the large $U$ regime as shown in Fig.\ref{Fig5}.

\begin{figure}[htb]
\centering
\includegraphics[width=8.0cm]{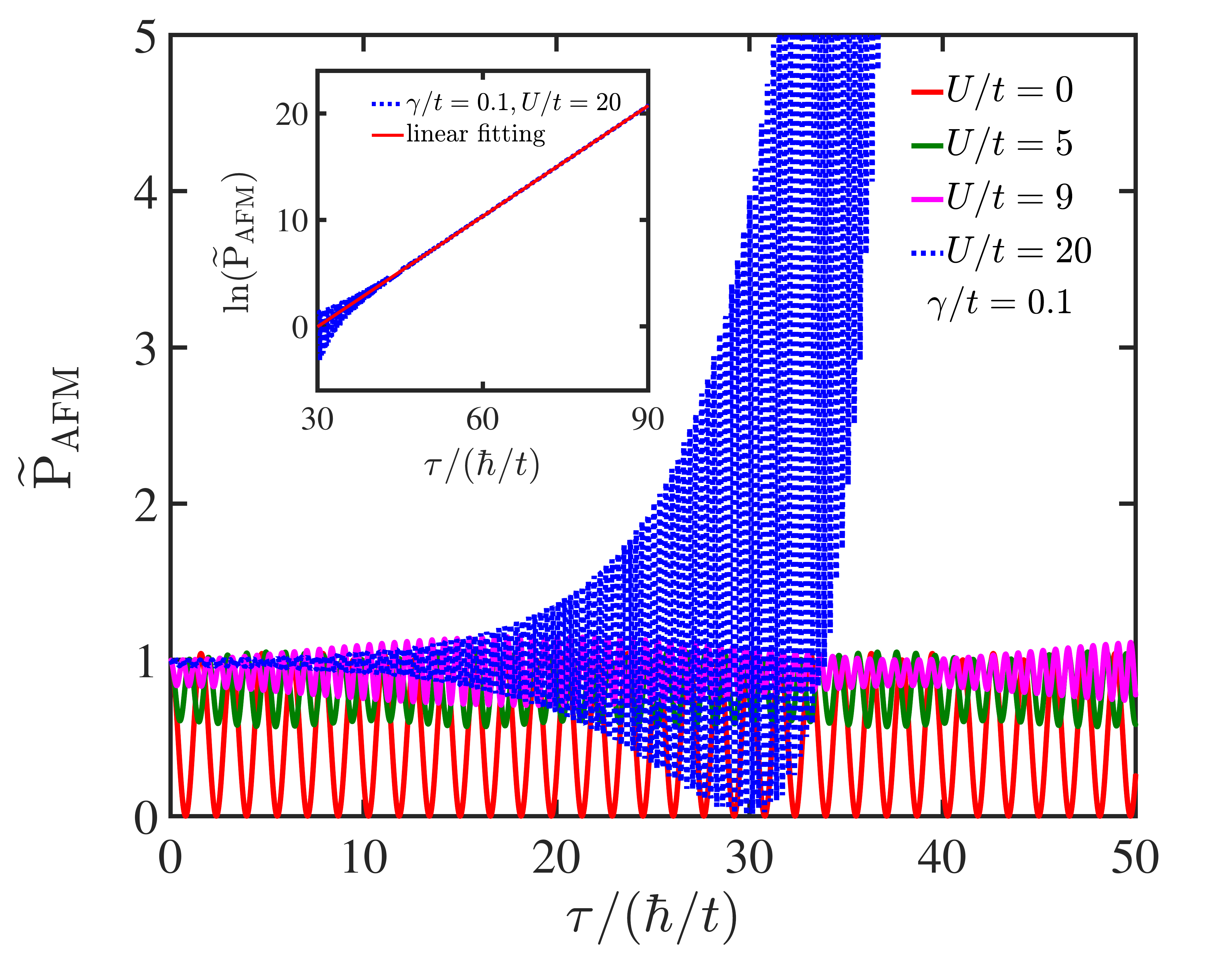}
\caption{Time evolutions of the rescaled probabilities of the AFM state in Mott phase $\widetilde{P}_{\rm AFM}(\tau)$ for different interaction strengths. In the inset, blue dotted line shows an accurate exponential growth which gives the lifetime $\tau/(\hbar/t)\approx30$ of Mott state. The solid line is the best fitting with an exponential function. Here the unit of time is $\hbar/t$.}
\label{Fig6}
\end{figure}

From Eq.(\ref{2.8}), we see the ground state is a Mott phase with AFM order, which seems stable as the ground state energy is always real even in the presence of dissipation. Nevertheless, the Mott phase is found to be dynamically unstable when the interaction strength $U$ exceeds a threshold.
For the system with initial state prepared in the AFM state, i.e., the ground state of Eq.(\ref{2.6}) or equivalently Eq.(\ref{2.8}) in the strong interaction regime, its dynamical properties are determined by all the eigenvalues and eigenvectors of $H_\mathcal{PT}$,  instead of Eq.(\ref{2.8}) solely. Therefore, the dynamical instability stems from the $\mathcal{PT}$ symmetry breaking in Eq.(\ref{2.9}). In the scheme of effective Hamiltonian, the rescaled probability can be calculated via
\begin{equation}
\begin{split}
\widetilde{P}_{\rm{AFM}}(\tau)&=\Big|\langle \mathbf{AFM}|e^{-iH_\mathcal{PT} \tau}|\mathbf{AFM}\rangle\Big|^2  \\
&=\Big|\sum_{n}\langle \mathbf{AFM}|\frac{e^{-iE_{n} \tau}|\psi_n^R\rangle \langle \psi_n^L|}{\langle \psi_n^L|\psi_n^R\rangle}|\mathbf{AFM}\rangle\Big|^2,
\end{split}
\label{2.10}
\end{equation}
where $|\psi_n^R\rangle$ and $|\psi_n^L\rangle$ are respectively the right and left eigenvector, which are defined via $H_{\mathcal{PT}}|\psi_{n}^{R}\rangle=E_{n}|\psi_{n}^{R}\rangle$,
$H_{\mathcal{PT}}^\dagger|\psi_{n}^{L}\rangle=E_{n}^*|\psi_{n}^{L}\rangle$ and
$\langle \psi_n^{{L,R}}|=\big(|\psi_{n}^{{L,R}}\rangle\big)^\dagger$. It can be found from Eq.(\ref{2.10}) that unstable dynamics is determined by combination of two factors: (positive) imaginary part of highly excited states and their overlap with AFM state. It follows from Eq.(\ref{2.9}) that the imaginary part is given by $2\mathbf{Im}\sqrt{\frac{t^4}{U^2}-\gamma^2}$ and the overlap between the highly excited states and AFM state is about $\frac{\sqrt{2}t}{U}$. Accordingly, we can define a timescale (lifetime) $\tau$ beyond which the rescaled probability of AFM state diverges exponentially.
The lifetime is estimated at $\tau \approx \frac{\ln\big(\frac{U}{\sqrt{2}t}\big)}{\mathbf{Im} \sqrt{\frac{t^4}{U^2}-\gamma^2}}$ and $\tau\approx30$ (in units of $\hbar/t$) for experimentally reachable parameters\cite{Bouganne} $\gamma/t=0.1$ and $U/t=20$, which matches well with the numerical result (see Fig.\ref{Fig6} and the inset).
Basically, the instability can be viewed as a signature of dynamical $\mathcal{PT}$-symmetry breaking which inevitably occurs along the time evolution even the nonzero overlap between AFM state and highly excited state with a finite imaginary part is small.

\section{Dissipative Fermi-Hubbard model: Multi-site system}\label{sec3}
In the preceding section, we have studied the non-Hermitian dynamics for two-site Fermi-Hubbard model.
In this section, we extend two-site to multi-site system and examine the non-Hermitian dynamics.
We consider the 1D Fermi-Hubbard model with balanced gain and loss on boundary, which is given by
\begin{equation}
H=H_s +i\gamma(n_L-n_1),
\label{3.1}
\end{equation}
where $H_s$ is the Hubbard Hamiltonian given by Eq.(\ref{Hubbard}) and the dissipative strength is denoted by $\gamma$. 
$n_1$ ($n_L$) denotes particle number on the first (last) site. It is obvious that the Hamiltonian (\ref{3.1}) is $\mathcal{PT}$ symmetric, i.e., $\mathcal{PT}H\mathcal{TP}=H$ according to the operations of $\mathcal{P}$ and $\mathcal{T}$
\begin{eqnarray}
\mathcal{P}c_j\mathcal{P}=c_{L+1-j},~~\mathcal{T}i\mathcal{T}=-i.\label{3.2}
\end{eqnarray}
Apart from an overall exponentially damping factor, we note that Eq.(\ref{3.1}) can be taken as an effective Hamiltonian of the dissipative Hubbard systems with fine-tune site-dependent dissipation parameters: $L_{k=(1,\sigma)}=\sqrt{6 \gamma} c_{1 \sigma}$, $L_{k=(i,\sigma)}=2 \sqrt{\gamma} c_{i \sigma}$ ($i=2,\cdots,L-1$), and $L_{k=(L,\sigma)}=\sqrt{2 \gamma} c_{L \sigma}$.

The Fermi-Hubbard model with spin-independent boundary fields can be exactly solved by Bethe ansatz \cite{BA1,BA2,BA3,BA4}. Here we generalize the real boundary fields to imaginary ones and derive the Bethe ansatz equations (BAEs) for Eq.(\ref{3.1}):
\begin{equation}
\begin{split}
&e^{i2k_j(L+1)}\frac{1-i\gamma e^{-ik_j}}{1-i\gamma e^{ik_j}}\frac{1+i\gamma e^{-ik_j}}{1+i\gamma e^{ik_j}}\\
&=\prod_{\beta=1}^{M}\frac{\sin k_j-\lambda_\beta+iu}{\sin k_j-\lambda_\beta-iu}\frac{\sin k_j+\lambda_\beta+iu}{\sin k_j+\lambda_\beta-iu},~~j=1,\ldots,N \\
& \prod_{l=1}^N\frac{\lambda_\alpha-\sin k_j+iu}{\lambda_\alpha-\sin k_j-iu}\frac{\lambda_\alpha+\sin k_j+iu}{\lambda_\alpha+\sin k_j-iu} \\
&=\prod_{\beta\neq\alpha}^{M}\frac{\lambda_\alpha-\lambda_\beta+2iu}{\lambda_\alpha-\lambda_\beta-2iu}
\frac{\lambda_\alpha+\lambda_\beta+2iu}{\lambda_\alpha+\lambda_\beta-2iu},~~\alpha=1,\ldots,M,
\label{BAEs}
\end{split}
\end{equation}
where $L$, $N$, $M$ denote the number of lattice sites, total particle number and particle number with spin down respectively, $u=\frac{U}{4t}$ and the energy eigenvalue is expressed by $E=-2t\sum_{j=1}^N\cos k_j$. The parameter sets $\{k_j\}$ denote the charge momenta and $\{\lambda_\alpha\}$ represent the spin rapidities which are introduced to describe the motion of spin waves. In strong interaction regime with $U/t \gg 1$ and $U/\gamma \gg 1$, the spin and charge degrees of freedom are separated and the quasi-momentum $k_j$ keep finite while $\lambda_\alpha$ are proportional to $u$. Expanding up to the first order with respect to $k_j/u$,
the ground state energy can be expressed by
\begin{eqnarray}
E=-\frac{t}{u}\frac{N}{L}\zeta,\label{3.3}
\end{eqnarray}
where $\zeta=\sum_{\alpha=1}^{M} \frac{4}{\Lambda_{\alpha}^{2}+1}$ with $\Lambda_{\alpha}=\lambda_\alpha/u$ and $\{\Lambda_{\alpha}\}$ satisfy the following equations
\begin{equation}
2N\theta\left(2\Lambda_{\alpha}\right)=2\pi J_{\alpha}+\sum_{\beta \neq \alpha}^{M}\left[\theta\left(\Lambda_{\alpha}-\Lambda_{\beta}\right)
+\theta\left(\Lambda_{\alpha}+\Lambda_{\beta}\right)\right],\label{3.4}
\end{equation}
where $\theta(x)=2\arctan(x/2)$ quantum number $J_{\alpha}$ are positive integers.
At the half-filling $N=L$, Eq.(\ref{3.4}) is nothing but the well-known BAEs of the open boundary Heisenberg model
\begin{equation}
H=J\sum_{i=1}^{L-1}\big(\mathbf{P}_{i, i+1}-1\big),\label{3.5}
\end{equation}
where $\mathbf{P}_{i, i+1}=\mathbf{S}_i\cdot \mathbf{S}_{i+1}+\frac{3}{4}$ is the permutation operator between the $i$th and $(i+1)$th spins.
Since the ground state energy of Hamiltonian (\ref{3.5}) with $J>0$ can be written as $E=-J\zeta$\cite{Sutherland}, by comparing it with Eq.(\ref{3.3}) we can find easily that the coupling constant $J=\frac{4t^2}{U}$. This is consistent with the result obtained by degenerate perturbation theory in which the half-filling Fermi-Hubbard is reduced effectively to anti-ferromagnetic Heisenberg model in single-occupied subspace when $U/t\gg1$.
\begin{figure}[htb]
\centering
\includegraphics[width=8.5cm]{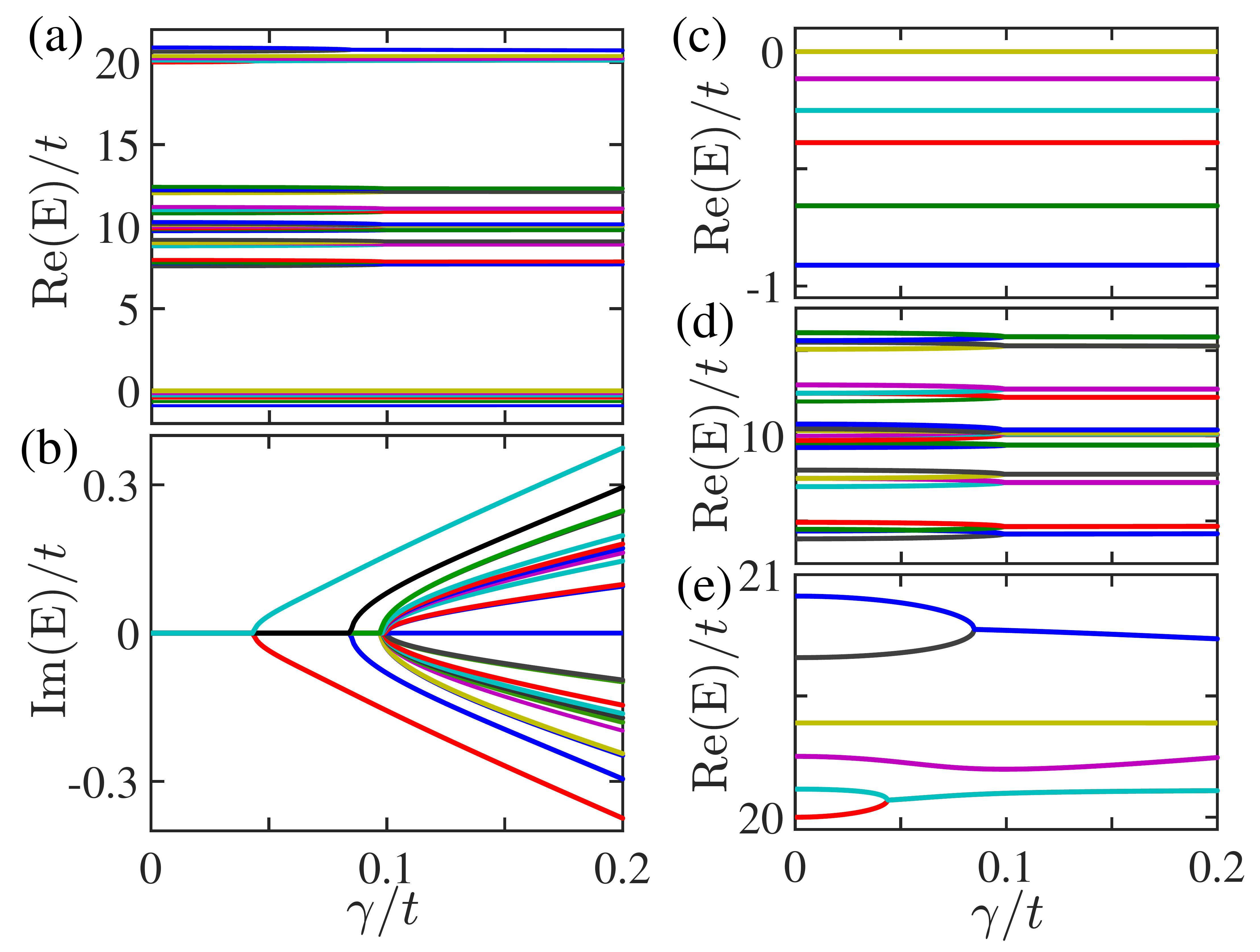}
\caption{Exact energy spectrum of four-site non-Hermitian Fermi-Hubbard model with $N_\uparrow=N_\downarrow=2$. (a) displays the real parts and (b) the imaginary parts versus $\gamma/t$ for  the system with $U/t=10$. (c), (d) and (e) are the enlarged figures of (a). The imaginary parts first appears at the highly excited states which consists mainly of two double-occupied states whose energy scale is about 20 as shown in (c) and then at the states with one double-occupied states with energy scale 10 (see (d)). The low-energy states always have real spectrum (see (e)).}
\label{Fig7}
\end{figure}

In Fig.\ref{Fig7}, we display the energy spectrum versus $\gamma/t$ for half-filling four sites system with balanced spin $N_\uparrow=N_\downarrow=2$ and $U=10$. As shown in Fig.\ref{Fig7} (a), the real parts of energy spectrum are divided into three regions, which are enlarged in Fig.\ref{Fig7} (c)-(d), respectively. We find that the spectrum of low-energy states
are always real for arbitrary $\gamma/t$ in the whole region shown in Fig.\ref{Fig7}(e), corresponding to single-occupied Mott states described by Heisenberg model.
As $\gamma$ increases, the complex conjugate pairs appear first in
two double-occupied states (Fig.\ref{Fig7}(c)) and then in one double-occupied states (Fig.\ref{Fig7}(d).
While all eigenvalues are real for small $\gamma/t$, the emergence of imaginary parts of energy eigenvalues in excited bands implies the breaking of $\mathcal{PT}$ symmetry when $\gamma/t$ exceeds a critical value. Similarly, if we fix $\gamma/t \ll 1$ and increase $U/t$, the complex eigenvalues will appear when $U$ exceeds a critical value.

\begin{figure}[htb]
\centering
\includegraphics[width=8.0cm]{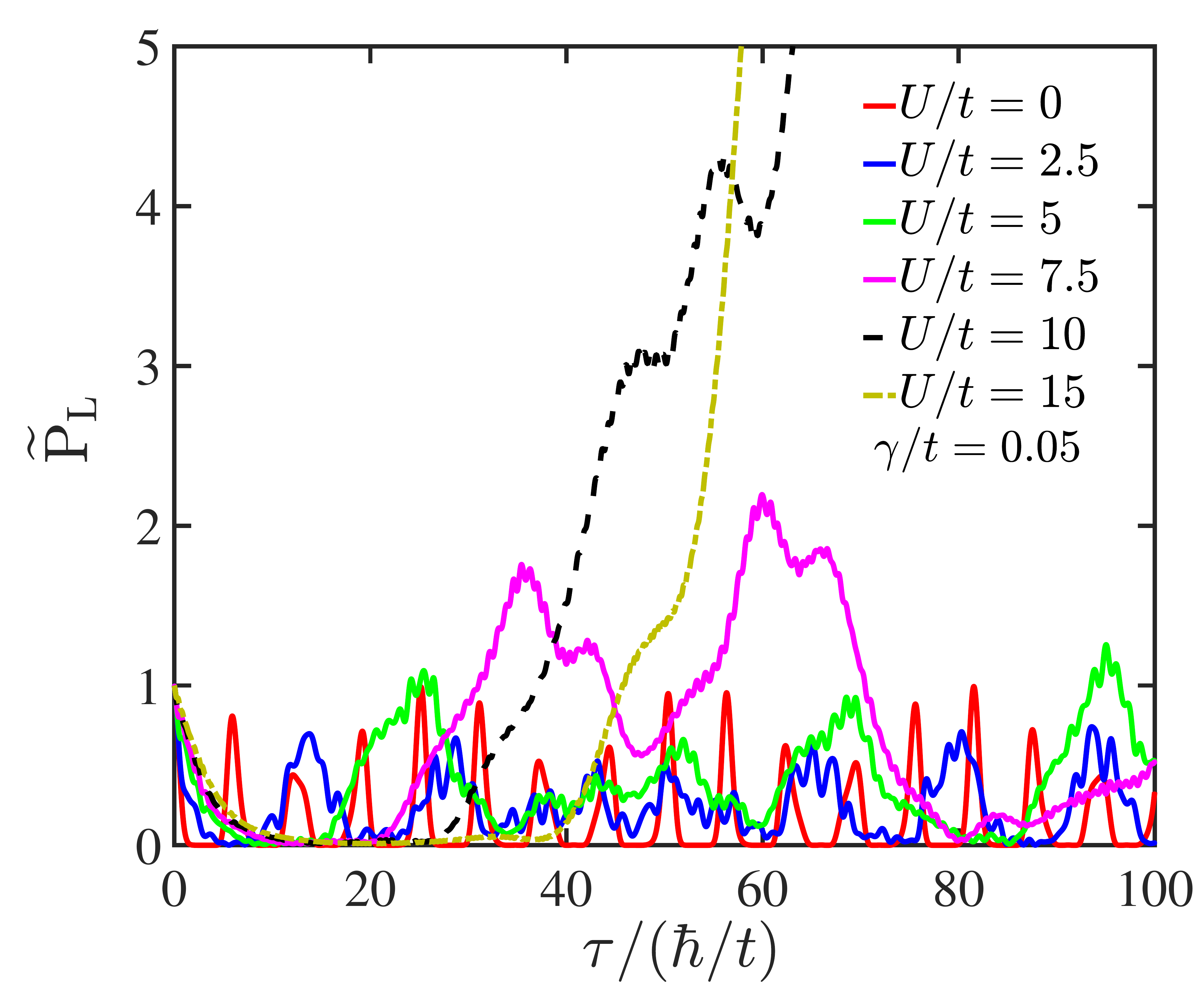}
\caption{Time evolutions of the rescaled probabilities of all four fermions occupying the left two sites $\widetilde{P}_L(\tau)$  of four-site system for different interaction strengths. Here $\gamma/t=0.05$ and the initial value is chosen by $\widetilde{P}_L(0)=1$. The unit of time is $\hbar/t$.}
\label{Fig8}
\end{figure}
\begin{figure}[htb]
\centering
\includegraphics[width=8.0cm]{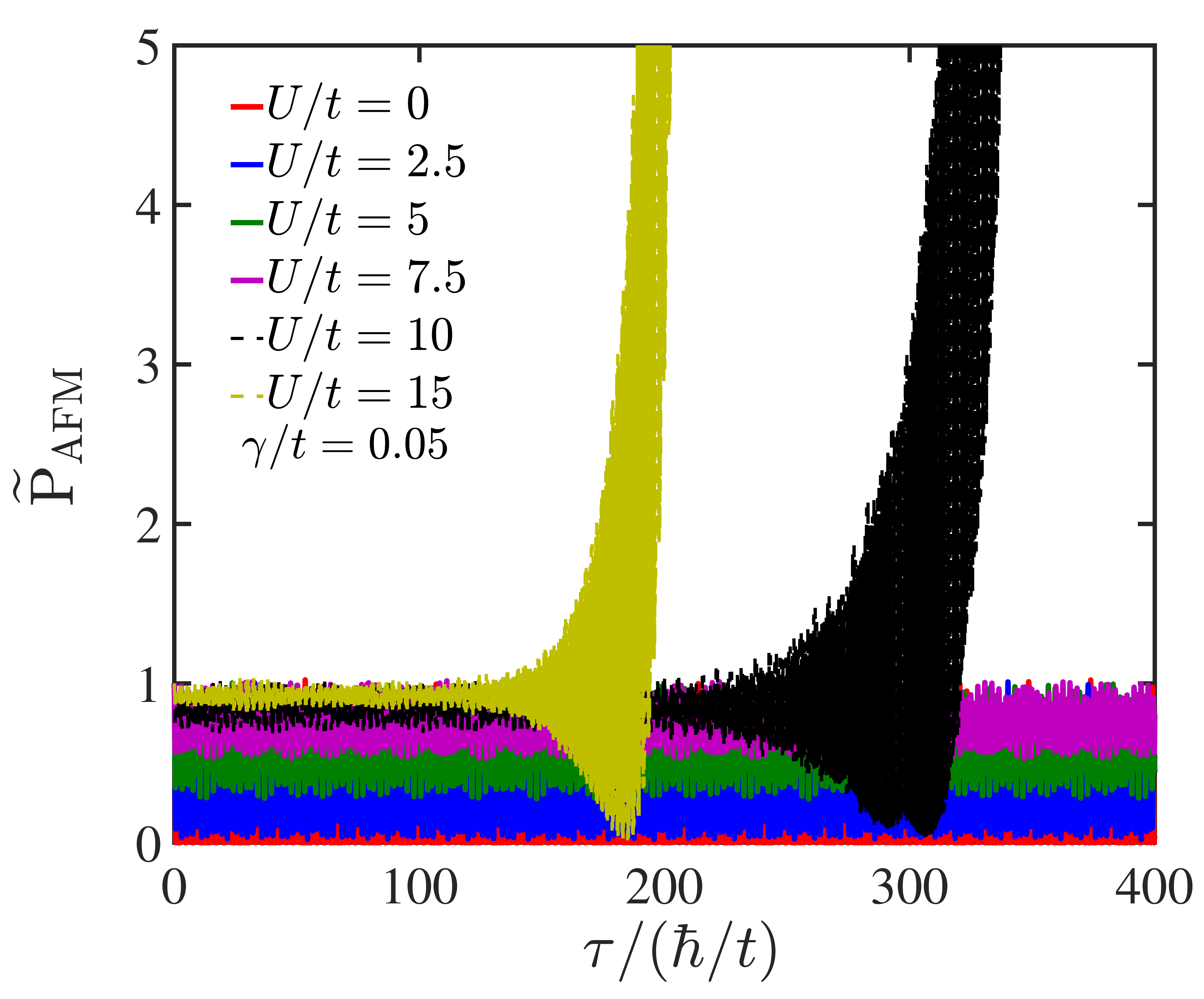}
\caption{Time evolutions of the rescaled probabilities of AFM (Mott) state $\widetilde{P}_{\rm AFM}(\tau)$ of four-site system for different interaction strengths which take the same values as in Fig.\ref{Fig8} . Here $\gamma/t=0.05$ and the initial state is chosen by $\widetilde{P}_{\rm AFM}(0)=1$. The unit of time is $\hbar/t$.}
\label{Fig9}
\end{figure}

In order to reveal the signature of dynamical $\mathcal{PT}$ symmetry breaking,  we define a rescaled density matrix as
\begin{eqnarray}
\widetilde{\rho}(\tau)\equiv e^{4 N\gamma \tau}\varrho_N( \tau).
\end{eqnarray}
Suppose that the initial state is prepared as the state with all four fermions occupying the left two sites,  we plot the rescaled probabilities of the state remaining in the initial state (rescaled return probability) in Fig.\ref{Fig8} for the system with $\gamma/t=0.05$ and various $U$. The interaction induced instability is clearly observed when the interaction exceeds a threshold, which is in qualitative agreement with results discussed in the double-well system. Similarly, for the dynamics of the AFM state (ground-state of Hamiltonian (\ref{3.5})), the interaction induced instability is also detected in the rescaled return probability as shown in Fig.\ref{Fig9}, which shares the same physical origin with the double-well system.
In order to estimate time for the emergence of dynamical instability in the rescaled return probability, we employ the same expression (\ref{2.10}) on this system. The characteristic time for dynamical $\mathcal{PT}$ symmetry breaking is given by
\begin{eqnarray}
\tau =\frac{\ln\big(1/|C|\big)}{\big|\mathbf{Im} E_{MI}\big|}, \label{3.6}
\end{eqnarray}
where $C$ denotes the overlap between the AFM (Mott) state and the double-occupied excited state whose energy $E_{MI}$ has maximum imaginary part.

Besides the model (\ref{3.1}), we note that some other effective Hamiltonians with $\mathcal{PT}$ symmetry can be also realized by engineering site-dependent dissipations. For example, if the dissipations are engineered only on the odd or even sites, i.e, $L_{k=(i,\sigma)}=2 \sqrt{\gamma} c_{i \sigma}$ with $i=2n-1$ or $2n$, apart from an overall damping term, we can get an effective Hamiltonian with $\mathcal{PT}$ symmetry, which reads as
\begin{equation}
H=H_s \pm \sum_{j=1}^{L}
(-1)^j i \gamma n_j ,
\end{equation}
where $H_s$ is the Hubbard Hamiltonian given by Eq.(\ref{Hubbard}) and $+$ ($-$) corresponds to the dissipations engineering on odd or even sites. Although the above Hamiltonians are no longer exactly solvable, they display similar physical phenomena as we studied in this work, i.e., the interplay of interaction and dissipation also induces $\mathcal{PT}$ symmetry breaking and leads to dynamical instability in the rescaled  return probability.

\section{Summary and outlook} \label{sec5}
In summary, we have studied the quantum dynamics in dissipative Fermi-Hubbard model with hidden dynamic $\mathcal{PT}$ symmetry breaking, which can be characterized by the emergence of dynamical instability in the evolution of rescaled density matrix. By studying the dissipative double-well system in detail, we demonstrate the equivalence of the Lindblad master equation and effective non-Hermitian Hamiltonian in dealing with the dynamical evolution of an initial state with fixed particle number. For the dissipation engineered only in one of the wells, we find that its effective Hamiltonian can possesses $\mathcal{PT}$ symmetry if we rescale the density operator, and the dynamical instability occurs when the interaction exceeds a critical value.  We further unveil that the effective Hamiltonian of multi-site Hubbard models with  site-dependent dissipations can possess $\mathcal{PT}$ symmetry and exhibit the interaction induced dynamical instability. This is originated from $\mathcal{PT}$ symmetry breaking of the effective Hamiltonian accompanied by the emergence of complex eigenvalues.  Particularly, for an effective Hubbard Hamiltonian with $\mathcal{PT}$ symmetry boundary fields, it is exactly solvable by applying the Bethe ansatz method. By analyzing the Bethe ansatz equations of multi-site systems, we find that the highly excited states are associated with complex eigenvalues, while the low-energy states described by effective AFM Heisenberg model always have real eigenvalues. However, the low-energy AFM state remains dynamically unstable due to the existence of $\mathcal{PT}$ symmetry breaking in the highly excited double-occupied bands.
Motivated by this work, it is valuable to explore more fascinating physical systems including both non-Hermitian effects and many-body correlations.

Finally, we discuss the experimental detection of the interaction induced dynamical $\mathcal{PT}$ symmetry breaking in dissipative optical lattices. For simplicity, we consider the double-well system in which the dissipation is engineered only at the left well. The experiment can initially prepare two fermions with one spin-up and one spin-down at the left well and at the same time apply a resonant laser to couple the lowest hyperfine state of atoms to their highly excited state $\ket{e}$, which generates the atom loss only at left well as shown in Fig.\ref{Fig1}. The initial state is $\sideset{_2}{}{\mathop{\bra{0}}}\leftidx{_1}{\bra{\downarrow\uparrow}}\varrho(\tau=0)\ket{\uparrow\downarrow}_1\ket{0}_2=1 $ and all other matrix elements equal to zero. Then, one can measure the atom number in the left (right) well at the time
$\tau$. The probability of finding both two fermions located at the left (right) well is given by
$p_L(\tau)=N_{\uparrow\downarrow}^L(\tau)/N(\tau)$ ($p_R(\tau)=N_{\uparrow\downarrow}^R(\tau)/N(\tau)$), where $N_{\uparrow\downarrow}^L(\tau)$ ($N_{\uparrow\downarrow}^R(\tau)$) denotes the measurement times of two fermions located at the left (right) well
at time $\tau$ and $N(\tau)$ is the total number of measurement times.
This probability can be calculated by $\sideset{_2}{}{\mathop{\bra{0}}}\leftidx{_1}{\bra{\downarrow\uparrow}}\varrho(\tau)\ket{\uparrow\downarrow}_1\ket{0}_2 $ ($\sideset{_2}{}{\mathop{\bra{\uparrow\downarrow}}}\sideset{_1}{}{\mathop{\bra{0}}}\varrho(\tau)\ket{0}_1\ket{\uparrow\downarrow}_2 $),
where $\varrho(\tau)$ can be obtained by solving the Lindblad master equation as shown in Sec.II. After the measurements, one can obtain the rescaled probability given by $\widetilde{P}_L(\tau)=e^{4 \gamma \tau}p_L(\tau)$ ($\widetilde{P}_R(\tau)=e^{4 \gamma \tau}p_R(\tau)$).
For a given dissipation strength $\gamma$, the interaction induced dynamical instability can be measured as the interaction strength increases to break the $\mathcal{PT}$ symmetry.
The same principle applies to the AFM Mott phase as long as the initial state is prepared in the AFM state \cite{Heidelberg,AFMHubbard,AFMHubbard2}. 
The anomalous dynamics revealed in this work is expected to be observable in the current Fermi-Hubbard experiments with dissipation\cite{DissHubbard}.
\\
\begin{acknowledgments}
We thank Z Cai for helpful conversation. The work is supported by NSFC under Grants No.11974413 and the National Key
Research and Development Program of China (2016YFA0300600 and
2016YFA0302104).
\end{acknowledgments}

~~~~~\\
~~~~~\\
~

\appendix

\section{Derivation of survival probability in double-well system}
In this appendix, we derive analytical solution of survival probability for two fermions occupying the left-well (Eq.(\ref{Eq8})) in main text. It can be written as 
\begin{widetext} 
\begin{equation}
\begin{split}
P_L(\tau)&=\sideset{_2}{}{\mathop{\bra{0}}}\leftidx{_1}{\bra{\downarrow\uparrow}}\varrho(\tau)\ket{\uparrow\downarrow}_1\ket{0}_2\\
&=\sideset{_2}{}{\mathop{\bra{0}}}\leftidx{_1}{\bra{\downarrow\uparrow}}e^{-i\mathcal{H} \tau}\varrho_N(0)e^{i\mathcal{H}^{\dagger} \tau}\ket{\uparrow\downarrow}_1\ket{0}_2
\\
&=\sideset{_2}{}{\mathop{\bra{0}}}\leftidx{_1}{\bra{\downarrow\uparrow}}\sum_j\frac{\left|\mu_{j}^{R}\right\rangle\left\langle\mu_{j}^{L}\right|}{\left\langle\mu_{j}^{L} | \mu_{j}^{R}\right\rangle}e^{-i\mathcal{H} \tau}\varrho_N(0)e^{i\mathcal{H}^{\dagger} \tau}\sum_k\frac{\left|\mu_{k}^{L}\right\rangle\left\langle\mu_{k}^{R}\right|}{\left\langle\mu_{k}^{R} | \mu_{k}^{L}\right\rangle}\ket{\uparrow\downarrow}_1\ket{0}_2
\\
&=\sideset{_2}{}{\mathop{\bra{0}}}\leftidx{_1}{\bra{\downarrow\uparrow}}\sum_j\frac{\left|\mu_{j}^{R}\right\rangle\left\langle\mu_{j}^{L}\right|}{\left\langle\mu_{j}^{L} | \mu_{j}^{R}\right\rangle}e^{-iE_j\tau}\varrho(0)\sum_ke^{iE_k^*\tau}\frac{\left|\mu_{k}^{L}\right\rangle\left\langle\mu_{k}^{R}\right|}{\left\langle\mu_{k}^{R} | \mu_{k}^{L}\right\rangle}\ket{\uparrow\downarrow}_1\ket{0}_2 
\label{P_L}
\end{split}
\end{equation}
\end{widetext} 
where $|\mu_{j}^{R}\rangle$ ($|\mu_{j}^{L}\rangle$) is the right (left) vector defined via $\mathcal{H}|\mu_{j}^{R}\rangle=E_{j}|\mu_{j}^{R}\rangle$ ($\mathcal{H}^{\dagger}|\mu_{j}^{L}\rangle=E_{j}^{*}|\mu_{j}^{L}\rangle$) and the identity  $\sum_{j}|\mu_{j}^{R}\rangle\langle\mu_{j}^{L}|/\langle\mu_{j}^{L}| \mu_{j}^{R}\rangle={\rm I}$ (where ${\rm I}$ is identity matrix)  is also considered.
Since the particle number $N$ is a good quantum number for the non-Hermitian effective Hamiltonian $\mathcal{H}$, then $\mathcal{H}$
can be written as a $4\times4$ matrix in the basis $\big\{\ket{\uparrow}_1\ket{\downarrow}_2, \ket{\downarrow}_1\ket{\uparrow}_2, \ket{\uparrow\downarrow}_1\ket{0}_2, \ket{0}_1\ket{\uparrow\downarrow}_2\big\}$ 
\begin{equation}        
\mathcal{H}=\left(                  
\begin{array}{cccc}    
0 & 0 & -t & -t\\   
0 & 0 & t & t\\   
-t & t & U-2i\gamma & 0\\   
-t & t & 0 & U+2i\gamma\\   
\end{array}
\right)                  
\end{equation}
and simple analytical expressions for its eigenvalues and eigenvectors can be obtained at the noninteracting limit of $U=0$. Substituting the $E_j,E_j^*$ and $|\mu_{j}^{L}\rangle$, $|\mu_{j}^{R}\rangle$ into equation (\ref{P_L}), one can obtain immediately
\begin{widetext} 
\begin{equation}
\begin{split}
P_L(\tau)=e^{-4 \gamma \tau}\left| \frac{-t^2 +2\gamma^2 \cosh(2\tau \sqrt{\gamma^2-t^2} ) -2 \gamma \sqrt{\gamma^2-t^2} \sinh(2\tau \sqrt{\gamma^2-t^2} )}{2 \gamma^2-t^2}\right|^2.
\end{split}
\end{equation}
\end{widetext} 
Thus, by the definition of $P_L(\tau)= e^{-4 \gamma \tau} f(\tau)$ and $\omega=\sqrt{\gamma^2-t^2}$, we arrive at Eq.(\ref{Eq8}) in main text. 

In fact, the above derivation is not dependent on the choice of initial state which means, in principle, one can study the survival probability for any quantum state. 

\end{document}